\documentstyle[12pt,dspp4]{article}

\newcommand{\ie}{i.e.,\ }
\newcommand{\eg}{e.g.,\ }
\newcommand{\etal}{et~al.\ }
\newcommand{\cf}{cf.\ }
\newcommand{\hubble}{\mbox{$H_0 = 100~h_{100}\,\rm km\, s^{-1}\, Mpc^{-1}$}}
\newcommand{\hmpc}{\mbox{$h_{100}^{-1}\, \rm Mpc$}}

\newcommand{\ihmpcc}{\mbox{$h_{100}^{3}\, \rm Mpc^{-3}$}}
\newcommand{\ihmpccmag}{\mbox{$h_{100}^{3}\, \rm Mpc^{-3}\, mag^{-1}$}}

\newcommand{\vhel}{\mbox{$v_{\rm hel}$}}

\newcommand{\vi}{\mbox{$V\!-\!I$}}
\newcommand{\magsq}{\mbox{$\rm mag\, arcsec^{-2}$}}

\newcommand{\mvvm}{\mbox{$\langle V/V_{max} \rangle$}}
\newcommand{\chisqnu}{\mbox{$\chi^{2}_{\nu}$}}

\begin{document}

\singlespace

\slugcomment{To appear in {\em The Astrophysical Journal}, Vol. 481,
June 1, 1997}

\bibliographystyle{apjaj40}


\title{Low Surface Brightness Galaxies in the Local Universe.
III. Implications for the Field Galaxy Luminosity Function}

\author{D.~Sprayberry}
\affil{Kapteyn Laboratorium, University of Groningen, Postbus 800,\\
9700 AV Groningen, The Netherlands\\
Email: dspray@astro.rug.nl}

\author{C.~D.~Impey}
\affil{Steward Observatory, University of Arizona, Tucson, AZ 85721\\
Email: cimpey@as.arizona.edu}

\author{M.~J.~Irwin}
\affil{Royal Greenwich Observatory, Madingley Road, Cambridge, UK CB3
0EZ\\
Email: mike@mail.ast.cam.ac.uk}

\and

\author{G.~D.~Bothun}
\affil{Department of Physics, University of Oregon, Eugene, OR 97403\\
Email: nuts@moo.uoregon.edu}

\begin{abstract}

We present a luminosity function for low surface brightness (LSB)
galaxies identified in the APM survey of \cite{impey96a}.  These
galaxies have central surface brightnesses ($\mu(0)$) in $B$ in the
range $22.0 \leq mu(0) \leq 25.0$.  Using standard maximum-likelihood
estimators, we determine that the best-fit Schechter function
parameters for this luminosity function (LF) are $\alpha = -1.42$,
$M^{*} = -18.34$, and $\phi^{*} = 0.0036$, assuming \hubble.  We
compare the luminosity and number densities derived from this
luminosity function to those obtained from other recent field galaxy
studies and find that surveys which do not take account of the
observation selection bias imposed by surface brightness are missing a
substantial fraction of the galaxies in the local universe.  Under our
most conservative estimates, our derivation of the LF for LSB galaxies
suggests that the CfA redshift survey has missed at least one third of
the local galaxy population.  This overlooked fraction is not enough
by itself to explain the large number of faint blue galaxies observed
at moderate redshift under no-evolution models, but it does help close
the gap between local and moderate-redshift galaxy counts.

\end{abstract}

\section{Introduction}

The optical luminosity function (LF) of galaxies is one of the
fundamental building blocks of cosmology.  Accurate knowledge of the
luminosity function is necessary for, among other things, estimating
the mean luminosity density of the universe, and predicting the
redshift distribution of objects in various magnitude intervals (see
\eg the review by Binggeli et~al.\ 1988\nocite{binggeli88}).
The shape of the luminosity function also provides an important test
for theories of galaxy formation (\eg Press \& Schechter
1974\nocite{press74}).  Further, considerable attention has been
focussed of late on the large numbers of blue galaxies found in deep
surveys, first described by \cite{kron80} and \cite{hall84}.  The
degree to which number counts of these galaxies exceed those predicted
from local observations (\eg Bruzual \& Kron 1980\nocite{bruzual80}
and Guiderdoni \& Rocca-Volmerange 1990\nocite{guiderdoni90}),
and indeed whether an excess exists at all (compare Koo et~al.\
1993\nocite{koo93} and McGaugh 1994\nocite{mcgaugh94b}), depend on the
shape, normalization and color dependence of the luminosity function.

One of the problems with building a galaxy luminosity function is that
surveys are limited in the detection of diffuse galaxies by the
brightness of the night sky, and in the detection of compact galaxies
by the difficulty in distinguishing stars and galaxies.  As
\cite{disney76} and \cite{disney83} have demonstrated, at a given
luminosity a survey will identify preferentially those galaxies that
have the maximum possible angular size above the limiting isophote.
At a constant luminosity, galaxies of high surface brightness (HSB)
become indistinguishable from stars, and galaxies of low surface
brightness (LSB) fall below the limiting isophote over most of their
extent.  Although they purport to be magnitude limited, galaxy surveys
which do not take account of surface brightness effects are missing an
unknown but potentially large number of galaxies in each magnitude
bin.  Recent surveys of the Virgo cluster by \cite{impey88} and of the
Fornax cluster by \cite{irwin90} and \cite{bothun91} have taken
account of this potential source of bias by deliberately searching for
LSB galaxies.  They have found that previous surveys missed a
significant fraction of the cluster populations, particularly at
fainter luminosities ($M_B \ga -16$), and \cite{impey88} determined
that inclusion of LSB galaxies in Virgo steepened the low-luminosity
tail of that cluster's luminosity function considerably.  To date,
however, no estimates of the field galaxy luminosity function have
addressed the effects of surface brightness bias.  However,
\cite{mcgaugh95a} found that the space density of galaxies as a
function of central surface brightness appears to be flat below
$\mu_B(0) = 22.0$.  Also, \cite{sprayberry96a} found a space density
of galaxies as a function of central surface brightness that appeared
flat below $\mu_B(0) = 23.0$ after descending from a peak around
$\mu_B(0) = 21.75$.  Although many of these LSB galaxies are not
necessarily faint, the forms of these distribution functions strongly
suggest that the normalization of the galaxy space density at $z=0$
has been strongly influenced by surface brightness selection effects.

We have recently completed a survey for LSB galaxies in the region
defined by $-3\deg \leq \delta \leq 3\deg$ and $|b| > 30\deg$,
surveying about 786 square degrees of sky with the Automated Plate
Measuring (APM) system at Cambridge.\footnote{The APM is a National
Astronomy Facility, at the Institute of Astronomy, operated by the
Royal Greenwich Observatory.  A general description of the APM
facility is given by \cite{kibblewhite84}.}  We have identified
693 galaxies, most previously uncataloged and most with central surface
brightness $\mu_B(0) > 22~\magsq$.  The complete catalog of this
survey appears in \cite{impey96a} (Paper I).  The selection effects
and completeness corrections for the survey are analyzed in detail in
\cite{sprayberry96a} (Paper~II).  

In this paper, we present the luminosity function for LSB galaxies
from the APM survey and compare that luminosity function to those
obtained from the CfA redshift survey.  We also review suggestions by
\cite{phillipps90}, \cite{mcgaugh94b}, \cite{mcleod94}, and
\cite{ferguson94a} that LSB galaxies might account at least partially
for the large numbers of faint blue galaxies seen in deep surveys.
Section~\ref{sec:smp} describes the survey data and presents the
samples used for determining the luminosity function and the
corrections applied to those samples.  Section~\ref{sec:mth} covers
the methods used to develop the luminosity functions.
Section~\ref{sec:res} presents the luminosity functions and compares
the results to those obtained from the CfA redshift survey.
Section~\ref{sec:imp} reviews the consequences of this LSB luminosity
function for the general field luminosity function and for the
question of local counterparts to the faint blue galaxies.  Finally,
Section~\ref{sec:cnc} summarizes our conclusions.  Throughout this
paper, we assume \hubble.  Also, all magnitudes and surface
brightnesses used here are in the Johnson $B$ band.

\section{Samples Used} \label{sec:smp}

The APM survey for LSB galaxies is presented in Paper~I, and Paper~II
describes the details of how LSB galaxies were identified and
calibrated.  Paper~II also presents a selection function that gives
the completeness of the survey as a function of galaxy central surface
brightness and scale length (hereafter, ``the APM selection
function'').

We conducted followup optical spectroscopy at the Multiple Mirror
Telescope\footnote{The Multiple Mirror Telescope is a facility jointly
operated by the Smithsonian Institution and the University of
Arizona.}  and 21 cm \ion{H}{1} spectroscopy at Arecibo
Observatory\footnote{The Arecibo Observatory is part of the National
Astronomy and Ionosphere Center.  The NAIC is operated by Cornell
University under a cooperative agreement with the National Science
Foundation.} to obtain radial velocities for as many of the
galaxies as possible.  To date we have measured recessional velocities
for 332 of the 693 galaxies on the list, of which 190 come from
\ion{H}{1} spectroscopy and 142 from optical spectroscopy.  These
heliocentric velocities are presented in Paper~I.  For developing the
luminosity function, we have further corrected these heliocentric
velocities to the rest frame of the Local Group, using the standard
correction $v_{corr} = \vhel + 300 \sin l \cos b$.  No correction was
applied for Virgocentric infall since the median velocity of the
sample places most of the galaxies well beyond the Local Supercluster.
These corrected velocities were then used to estimate distance moduli
using the relation:
\begin{equation}
m - M = 5\left[\log v_{corr} - \log H_0 + 5\right]
\label{eqn:dmod}
\end{equation}
assuming as noted above that \hubble.  

\begin{figure*}[t]
\epsfbox{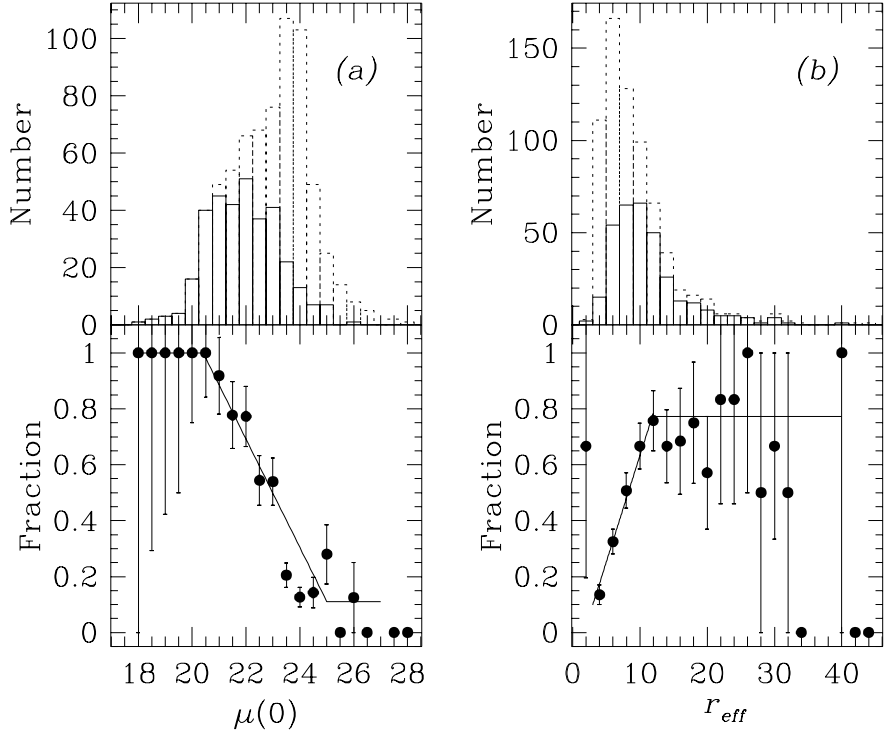}

\caption[Structural properties of complete LSB sample and subset with
velocities]{Structural properties of the complete LSB sample and the
subset with radial velocities.  {\em (a)} shows the distribution as a
function of $B$ central surface brightness, and {\em (b)} shows the
distribution as a function of half-light radius.  In the upper panels,
the dotted histogram is the distribution of the complete sample, and
the solid histogram is the distribution of the subset with velocities.
In the lower panels, the filled circles show the fraction of galaxies
with velocities for each bin, with error bars from counting
statistics.  The solid lines show the parametrizations described in
the text. \label{fig:asobs}}

\end{figure*}

The galaxies with velocities do not form a random subset of the
overall survey.  For reasons of observational efficiency, like all
other galaxy surveyers we favored galaxies of higher central surface
brightness and larger angular size.  Figure~\ref{fig:asobs} shows
the distributions of central surface brightness and half-light radius
for the complete sample and for the subset with velocities, along with
the ratios of the two sets by bin.  We assume that the galaxies for
which we have measured redshifts are representative of all galaxies in
a given bin of surface brightness and angular size.  This additional
source of bias must be taken into account in preparing a luminosity
function.  We have parameterized this bias in the simple forms
depicted in Figure~\ref{fig:asobs}: three separate linear fits in
the different regions of the $\mu(0)$ distribution
\begin{equation}
p_{\mu} = \left\{ \begin{array}{r@{,\quad}l}
	1.000 & \mu(0) < 20.25 \\
	4.950 - 0.194\,\mu(0) & 20.25\leq \mu(0) \leq 25.0 \\
	0.111 & \mu(0) > 25.0 \end{array} \right.
\label{eqn:muasobs}
\end{equation}
where $\mu(0)$ is in \magsq, and in the different regions of the 
half-light radius distribution
\begin{equation}
p_{r_e} = \left\{ \begin{array}{r@{,\quad}l}
	0.667 & r_{eff} < 3 \\ 
	-0.130 + 0.076\,r_{eff} & 3\leq r_{eff} 
		\leq 13\\
	0.773 & r_{eff} > 13 \end{array} \right.
\label{eqn:reasobs}
\end{equation}
where $r_{eff}$ is in arcseconds.  The final probability that an LSB
galaxy will be detected by the APM and
included in the subset with velocities is given by
\begin{equation}
p_{tot} = p_{APM} \times p_{\mu} \times p_{r_e}
\label{eqn:correct}
\end{equation}
where $p_{APM}$ is the probability derived from the APM selection
function of Paper~II.  Equation~\ref{eqn:correct} assumes that the
corrections in $\mu(0)$ and $r_{eff}$ are separable.  This assumption
is reasonable for our sample, because $\mu(0)$ and $r_{eff}$ are
uncorrelated: Pearson's $r = 0.075$ and the Spearman rank correlation
coefficient $s = -0.121$, and neither coefficient is significantly
different from zero.

We note that Figure~\ref{fig:asobs} shows that the surface brightness
range $23.5 \le \mu(0) \le 24.5$ includes a large number of identified
galaxies, but that a very small fraction of those galaxies were
observed spectroscopically.  Also, the observed fraction as a function
of angular size declines sharply at small sizes.  These features are
artifacts of the two stages in which the APM survey was performed.  The
first stage identified LSB galaxies of large angular size, and all the
followup spectroscopy was performed on galaxies in this first list.
The second stage identified small angular size galaxies, which also
tended to be predominantly in the surface brightness range $23.5 \le
\mu(0) \le 24.5$.  The interested reader is referred to Paper~II for a
more complete discussion of the survey mechanics.  Here we note only
that the actual observed fraction in the range $23.5 \le \mu(0) \le
24.5$ lies {\em below} the parametrization of
Equation~\ref{eqn:muasobs}, which implies that the parametrized
correction is too small for those two surface brightness bins.  Any
bias introduced by this effect is ``conservative'', in that it will
result in an underestimation of the total number of LSB galaxies.  

We can estimate the completeness of our sample of galaxies using the
\mvvm\ test of \cite{schmidt68}.  For the complete set of 693 galaxies
identified by the APM, the test yields $\mvvm = 0.15\pm 0.04$ with no
corrections for incompleteness, and $\mvvm = 0.44\pm 0.06$ after
correcting for incompleteness using the APM selection function
described in Paper~II.  For the subset of 332 galaxies with
velocities, the test gives $\mvvm = 0.04\pm 0.05$ with no corrections
for incompleteness, $\mvvm = 0.34\pm 0.07$ after applying just the APM
selection function, and $\mvvm = 0.50\pm 0.07$ after applying the APM
selection function and the further correction for incompleteness in
the velocity observations from Equations~\ref{eqn:muasobs},
\ref{eqn:reasobs}, and \ref{eqn:correct} (as depicted in
Figure~\ref{fig:asobs}). The corrections thus substantially remove the
incompleteness in both the complete set and in the subset chosen
for spectroscopy.

There is yet another source of bias to be found in the magnitudes
measured for LSB galaxies.  The magnitudes measured in our survey are
isophotal magnitudes, not extrapolated or asymptotic.  The median
limiting isophote is $\mu_{lim} \approx 27.4$ \magsq.  As authors
from \cite{disney76} to \cite{mcgaugh94b} have pointed out, use of
isophotal magnitudes will cause galaxy luminosities to be
underestimated, and the underestimation becomes more severe with
decreasing central surface brightness.  Most LSB galaxies are
well-described by exponential surface brightness profiles (Impey
et~al.\ 1988\nocite{impey88}, Bothun et~al.\ 1991\nocite{bothun91},
and McGaugh \& Bothun 1994\nocite{mcgaugh94}) of the form
\begin{equation}
\mu(r) = \mu(0) + 1.086\frac{r}{l}
\label{eqn:exprof}
\end{equation}
where $\mu(0)$ is the central surface brightness in \magsq\ and $l$ is
the exponential scale length in arcseconds.  This simple analytical form allows a
direct calculation of the ratio of the total galaxy flux to that
observed within the limiting isophote, as
\begin{equation}
\frac{F_{obs}}{F_{tot}} = 1 - (1+n_{l})e^{-n_{l}}
\label{eqn:fobs}
\end{equation}
where $n_{l}$ is the number of scale lengths $l$ observed within the
limiting isophote.  This simple approximation will clearly understate
the ratio for galaxies with central condensations, such as spirals
with bulges.  The isophotal aperture in units of the galaxy scale
length is then given by
\begin{equation}
n_{l} = \frac{\mu_{lim} - \mu(0) - 10 \log (1+z) -k(z)}{1.086}
\label{eqn:apsize}
\end{equation}
where $\mu_{lim}$ is the surface brightness of the limiting isophote.
The first term involving $z$ accounts for the $(1+z)^4$ cosmological
dimming in surface brightness, and the second corrects for the
redshifting of the galaxy's spectral energy distribution (the $k$
correction).  The $k$ correction of course depends on galaxy type as
well as redshift.  The magnitudes and surface brightness for the LSB
galaxies with velocities have been corrected as described in
Paper~II using the tabulated $k$ corrections of
\cite{coleman80}.  The \bv\ and \vr\ colors for galaxy types Sbc,
Scd, and Irr closely match the range of colors observed among the 
galaxies for which we obtained CCD photometry.  The absolute
magnitudes have been corrected according to
Equations~\ref{eqn:apsize} and \ref{eqn:fobs}, so as to avoid 
skewing the luminosity function by this tendency to underestimate 
galaxy luminosities.  

Of course, our set of LSB galaxies is not itself a fair sample of the
local galaxy population, precisely because it excludes most galaxies
with $\mu(0) \la 22~\magsq$.  However, it is still useful to derive a
luminosity function for this set, so that this LF can be compared to
one derived from higher surface brightness galaxies.  In this way, it
is possible to obtain some idea of how surface brightness selection
effects have influenced estimates of the density of local galaxies
(see also McGaugh et~al.\ 1995\nocite{mcgaugh95a} and Paper~II).  To
validate such a comparison, it is necessary first to compare the range
of surface brightnesses covered by the present set of LSB galaxies
with the range covered by other surveys.  Unfortunately, no other
recent galaxy redshift surveys have published surface brightness data
for their galaxies.  Thanks to the recent release of a digitized
version of the original Palomar Observatory Sky Survey (the
Digitized Sky Survey\footnote{Based on photographic data of the
National Geographic Society -- Palomar Observatory Sky Survey
(NGS-POSS) obtained using the Oschin Telescope on Palomar Mountain.
The NGS-POSS was funded by a grant from the National Geographic
Society to the California Institute of Technology.  The plates were
processed into the present compressed digital form with their
permission.  The Digitized Sky Survey was produced at the Space
Telescope Science Institute under US Government grant NAG W-2166.} or
DSS), it is now possible to make independent measurements of the basic
photometric parameters of any object visible on the original survey,
when the celestial coordinates of the surveyed galaxies are known.
The CfA Redshift Survey described by \eg \cite{marzke94} is
based on Zwicky's Catalog of Galaxies and Clusters of Galaxies, which
was in turn created by visual examination of the Palomar Observatory
Sky Survey plates, so every object included in that survey should be
visible on the DSS.  Most importantly, the coordinates of galaxies
surveyed by the CfA are publicly available, so that it is possible to
retrieve images of the surveyed galaxies from the DSS.  Thus it should
be possible to measure the surface brightness range covered by the CfA
Redshift survey.  The lack of publicly available coordinates prevents
us from making a similar analysis of other recent redshift surveys.

We recovered from the Astrophysics Data System listing of the CfA
Redshift Survey the coordinates of every galaxy listed in the regions
of sky used by \cite{marzke94}.  We subdivided that list according to
the morphological categories used by \cite{marzke94a}, and we randomly
selected 10\% of the galaxies within each morphological class to keep
the number of galaxies manageable.  This selection yielded a list of
579 galaxies.  We then retrieved images from the DSS of this randomly
chosen subset and analyzed the images using the same algorithms used
in our APM LSB galaxy survey.  In this way, we obtained extrapolated
central surface brightnesses for the CfA galaxies that are directly
comparable to those obtained in the course of the APM survey.
Paper~II contains a complete description of the process of estimating
the extrapolated central surface brightness.  As a check on the
calibrations, we also retrieved from the DSS images of a randomly
chosen subset of the APM LSB galaxies and analyzed them.  After
cross-calibration, the results for the APM LSB galaxies were
consistent with those obtained from the deeper UKST plate materials
used in the APM LSB survey, with the exception that the lowest surface
brightness objects were not visible on the DSS.

\begin{figure*}[t]
\epsfbox{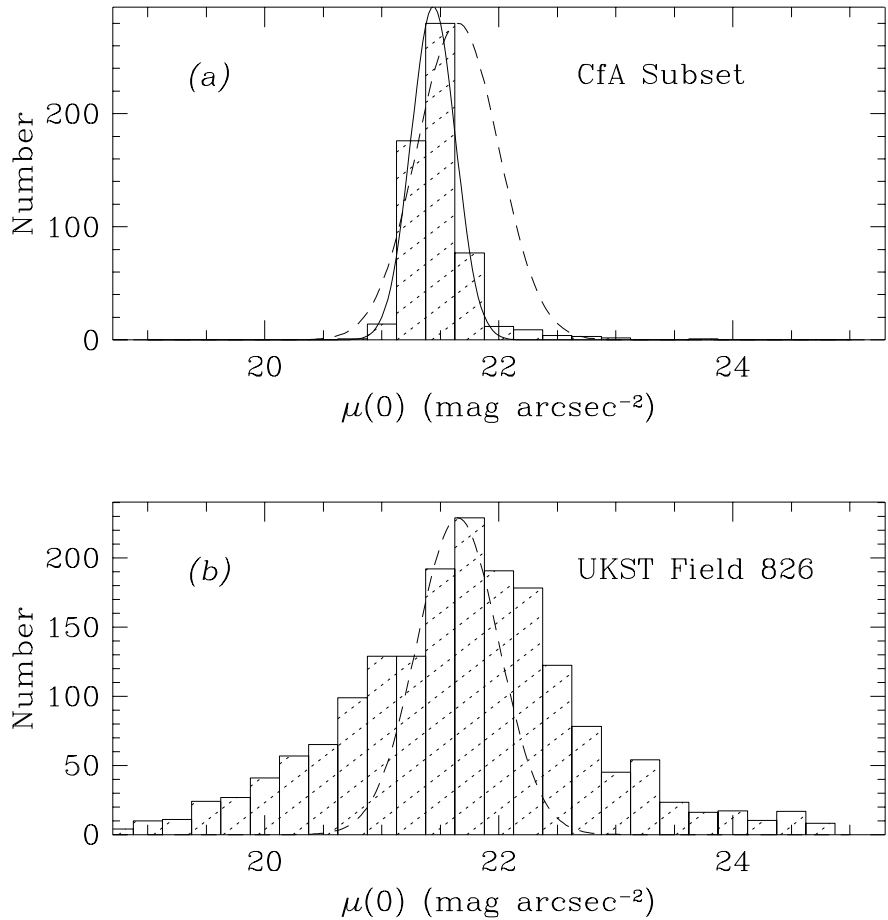}

\caption[Distributions of central surface brightness]{Distributions of
central surface brightness for {\em (a)} a randomly chosen sample of
galaxies from the CfA redshift survey, and {\em (b)} the complete list
of galaxies identified by machine scan of one UKST survey field.  In
{\em (a)} the solid curve represents the best fit of a Gaussian to the
CfA survey surface brightness distribution.  In both panels the dashed
Gaussians illustrate the canonical ``Freeman Law'' of $\mu (0) = 21.65
\pm 0.35$.  The distribution in (b) is corrected for incompleteness of
the detection algorithm for $\mu_B(0) \la 25$ as described in
Paper~II.  \label{fig:sbhists} }

\end{figure*}

\begin{figure*}[t]
\epsfbox{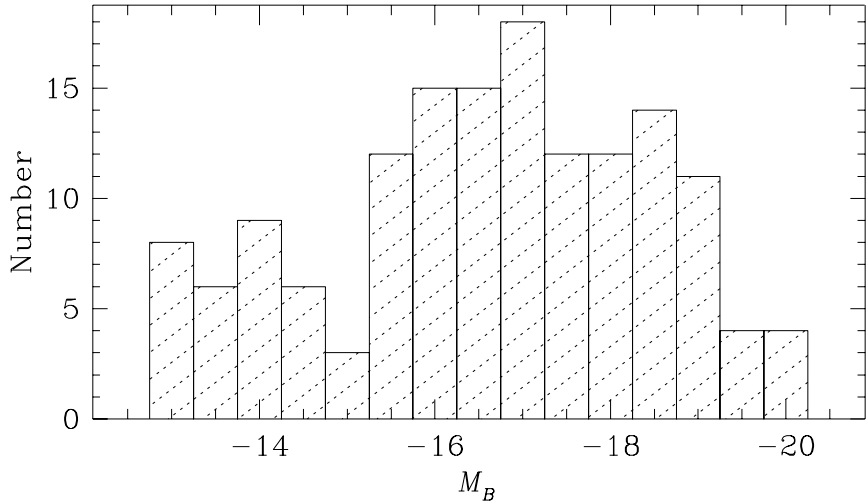}

\caption[Distribution of absolute magnitudes]{Distribution of absolute
magnitudes for the LSB galaxies ($\mu(0) > 22.0$ \magsq) used to
develop the LF.  This distribution includes the effects of the
correction from isophotal to total magnitudes described in
Equations~\ref{eqn:apsize} and \ref{eqn:fobs}.\label{fig:absmaghist}}

\end{figure*}

The surface brightness distribution for the CfA Redshift survey is
shown in the upper panel of Figure~\ref{fig:sbhists}.  The solidly
drawn smooth curve represents the best Gaussian fit to the CfA
distribution.  The lower panel shows the complete SB distribution
obtained by the APM for one UKST field.  Also drawn for illustration
in each panel is a dashed curve representing the canonical ``Freeman
Law'', a Gaussian centered at $\mu (0) = 21.65$ with $\sigma = 0.35$
(Freeman~1970\nocite{freeman70}).  It is clear from
Figure~\ref{fig:sbhists} that the range of surface brightnesses
covered by the CfA Redshift Survey is very narrow, narrower even than
the ``Freeman Law.''  The best-fit Gaussian to the CfA distribution
has a center at $\mu(0) = 21.44$ and $\sigma = 0.19$.  This is
completely consistent with the investigation of the Zwicky magnitude
scale by \cite{bothun90a} who find that this magnitude is not a
sky-limited magnitude.  In this case, one expects surface brightness
effects to completely dominate the magnitude estimates.  In essence,
the Zwicky magnitude is very much a ``bulge'' or high surface
brightness magnitude and is insensitive to extended, low surface
brightness light.  In contrast, the APM LSB survey has identified
galaxies over a much broader range, as described in Paper~II.
Clearly, the identification of galaxies for the CfA Redshift survey
suffered from a substantial bias against LSB galaxies.  In all the
following analysis, we use only those galaxies from the APM LSB survey
with $\mu(0) > 22.0\, \magsq$, or $3\sigma$ fainter than the typical
value found in the CfA Survey.  This limitation assures that the
resulting LF covers a different regime of surface brightness parameter
space from that covered by the LFs of \cite{marzke94} and
\cite{marzke94a}.  We note that there is a weak LSB tail in the CfA
distribution: the overall \chisqnu\ of the Gaussian fit is 1.37,
virtually all of which is due to this tail.  However, the very
weakness of this tail, when compared to the APM distribution in the
lower panel, underscores the severity of the SB selection bias
inherent in the CfA survey.  We note also that the CfA survey does not
identify nearly as many high surface brightness galaxies as does the
APM.  This lack is most likely due to the general absence of galaxies
smaller than 1 arcminute from the Zwicky catalog; many of the high
surface brightness galaxies identified by the APM are smaller than 1
arcminute.  Figure~\ref{fig:absmaghist} shows the distribution of
absolute magnitudes for the LSB survey galaxies with $\mu(0) > 22.0\,
\magsq$.

\section{Methods} \label{sec:mth}

The differential luminosity function of field galaxies $\phi(M)dM$ is
defined as the function giving, at each absolute magnitude $M$, the
number of galaxies per $\rm Mpc^{-3}$ in the luminosity interval $M +
dM/2 \leq M \leq M - dM/2$.  Because the area surveyed by the APM LSB
survey covers a wide area of sky and cuts across several large scale
structures, we adopt two density-independent techniques for estimating
the LF.  The first is the parametric maximum likelihood technique
developed by \cite{sandage79} (hereafter STY).  The second is the
stepwise maximum likelihood method (hereafter SWML) developed by
\cite{efstathiou88}.  Both methods assume that the LF has a universal
form, independent of position, allowing the probability of a galaxy's
inclusion in a complete catalog to be written simply in terms of the
LF itself.  The STY method is continuous and uses all the galaxy data,
but it requires the assumption of a parametrized form for the LF.  It
therefore gives no information as to the suitability of the
parametrized form chosen to represent the LF.  The SWML method
requires binning the data, but it requires no assumptions about the
shape of the LF.  It can therefore be used in combination with the STY
method to provide an independent check on the goodness-of-fit of the
chosen parametrization, as described by \cite{efstathiou88}.  Like
\cite{marzke94a} and virtually all others who have used this
combination of methods, we assume in the STY method a luminosity
function parameterization in the form first proposed by
\cite{schechter76}, which is written in absolute magnitudes as
\begin{equation}
\phi (M)dM = 0.4 \ln 10 \phi_{*} \left[ (10^{0.4(M_{*} - M)})^{1+\alpha}
e^{-(10^{0.4(M_{*} - M)})} \right] dM
\label{eqn:schechter}
\end{equation}
Using the two methods together thus gives best-fit values for the
Schechter function parameters $\alpha$ (the faint-end slope) and
$M^{*}$ (the characteristic absolute magnitude of the ``knee''), as
well as a probability that the underlying galaxy population is
well-described by the best-fit Schechter function.  

There is one major difficulty with applying these methods to the APM
LSB galaxy survey data.  Both the STY method and the SWML method
assume that the galaxy catalog in use is magnitude limited, or that
all galaxies with $m < m_{lim}$ have the same probability ($p < 1$) of
being included in the catalog, as in the case of a redshift survey
that uniformly samples a magnitude-limited catalog with $1/n$
sampling.  In our case, however, each galaxy has a unique probability
of inclusion that is determined from Equation~\ref{eqn:correct}, so
the given forms of the STY and SWML methods require modification.
\cite{zucca94} recently addressed this problem.  They derived a simple
modification to the STY estimator that accounts for the unique
observation probability assigned to each galaxy:
\begin{equation}
{\cal L} = \prod_{i=1}^N p_i^{w_i}
\label{eqn:zuccasty}
\end{equation}
where $\cal L$ is the likelihood to be maximized, the weight $w_i$ is
defined as the inverse of the probability that the $i$th galaxy will
be included in the sample (\ie for our situation $w_i =
1/p_{tot,i}$, with $p_{tot,i}$ from Equation~\ref{eqn:correct}),
and $p_i$ is as defined by STY:
\begin{equation}
p_i = \phi (M_i) \left/ \int_{M_{max(z_i)}}^{- \infty}
\phi (M)  dM \right.
\label{eqn:styprob}
\end{equation}
The corresponding change to the SWML
estimator of \cite{efstathiou88} immediately yields:
\begin{equation}
\ln {\cal L} = \sum_{i=1}^N W(M_i - M_k) w_i \ln \phi_k -
\sum_{i=1}^N w_i \ln \left\{ \sum_{j=1}^{N_p} \phi_j \Delta M
H\left(M_{max(z_i)} - M_j\right) \right\} + {\rm const}
\label{eqn:zuccastep}
\end{equation}
where the $\phi_k$ are the luminosity function values within each bin,
$N$ is the total number of galaxies in the sample, $N_p$ is the number
of steps, $M_{max(z_i)}$ is the maximum (\ie the faintest)
absolute magnitude visible at $z_i$, $\Delta M$ is the bin width in
magnitudes, and the window functions are
\begin{equation}
W(x) = \left\{ \begin{array}{r@{,\quad}l}
        1 & |x| \leq \Delta M/2 \\ 0 & {\rm otherwise}
        \end{array} \right.
\label{eqn:www}
\end{equation}
and
\begin{equation}
H(x) = \left\{ \begin{array}{r@{,\quad}l}
        0 & x < -\Delta M/2 \\
        (x/\Delta M + 1/2) & |x| \leq \Delta M/2 \\
        1 & x > \Delta M/2
        \end{array} \right.
\label{eqn:hhh}
\end{equation}
There is an implied sum over the doubled index $k$ in the first term
of Equation~\ref{eqn:zuccastep}. 

Finally, the survey biases must also be incorporated into the
normalization.  Both the STY and SWML estimators are normalized in the
manner described by \cite{efstathiou88} using the unbiased minimum
variance estimate of the mean density as developed by \cite{davis82},
but with a modification to the estimator to incorporate the
corrections for survey incompleteness.  This normalization proceeds in
three steps.  First, a selection function is defined as
\begin{equation}
S(x) = \int_{max\left[M_{max(x)},M_2\right]}^{M_1} \phi (M) dM \left/ %
\int_{M_2}^{M_1} \phi (M) dM \right.
\label{eqn:normprob}
\end{equation}
for galaxies in the range $M_1 < M < M_2$, where $M_{max(x)}$ is the
maximum (\ie the faintest) absolute magnitude visible at
distance $x$ according to the catalog limits.  Second, this selection
function is then corrected to incorporate the incompleteness
correction, so that it includes the combined probability of detecting
and spectroscopically observing an LSB galaxy in our survey:
\begin{equation}
S_{tot}(x_i) = S(x_i) \times p_{tot,i}
\label{eqn:normlsb}
\end{equation}
where $p_{tot,i}$ is obtained from \ref{eqn:correct}.  Finally, the
mean density of galaxies is obtained from the corrected selection
function as described by \cite{efstathiou88}:
\begin{equation}
\langle n\rangle = \frac{1}{V}\sum_{i=1}^N \frac{1}{S_{tot}(x_i)}
\label{eqn:meandens}
\end{equation}
where the sum extends over all the galaxies in volume $V$.  The mean
density is converted to a Schechter function normalization as:
\begin{equation}
\phi_{*} = \frac{\langle n \rangle}{\Gamma\left(\alpha +1,
10^{0.4(M_{*} - M_2)}\right) - \Gamma\left(\alpha +1,
10^{0.4(M_{*} - M_1)}\right)}
\label{eqn:phistar}
\end{equation}
where $\Gamma$ is the Euler incomplete gamma function.

\cite{zucca94} also estimated the effects of failing to consider
the individual galaxy weights.  Their simulations revealed that use of
Equation~\ref{eqn:styprob} to determine the Schechter function
parameters for a galaxy sample with significant incompleteness ($\mvvm
\la 0.3$) would bias the results towards flatter faint-end slopes
(\ie lower absolute values of $\alpha$) and
brighter values of $M_{*}$.  We can objectively determine individual
galaxy weights from parameters of our survey technique (the APM
selection function) and from the internal statistics of our followup
observations (Figure~\ref{fig:asobs} and
Equations~\ref{eqn:muasobs} and
\ref{eqn:reasobs}), so Equations~\ref{eqn:zuccasty} and
\ref{eqn:zuccastep} are the clear techniques of choice for our
data.

\section{Results} \label{sec:res}

\begin{figure*}[t]
\epsfbox{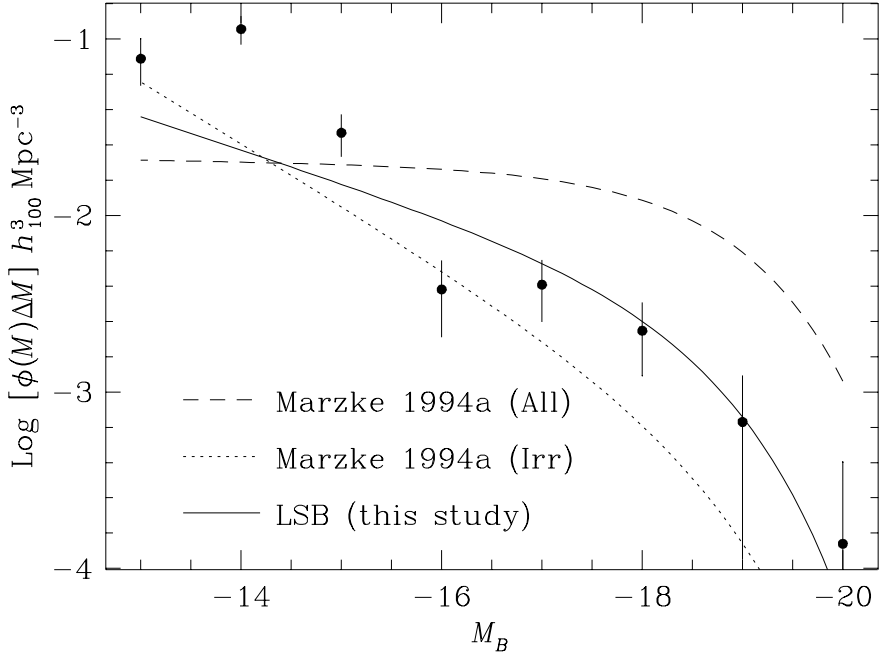}

\caption[Luminosity function for LSB galaxies]{Luminosity function for
LSB galaxies from the APM survey.  The solid line represents the
maximum likelihood Schechter function, and the points with error bars
represent the model-independent step-wise maximum likelihood function.
Note that for the LSB galaxies, the model-independent binned LF shows
a significant excess of low luminosity galaxies, beyond the level of
the maximum likelihood Schechter function.  The dashed line shows the
maximum likelihood Schechter function estimated by \cite{marzke94a}
for all morphological types in the CfA redshift survey, and the dotted
line shows the maximum likelihood Schechter function estimated by
Marzke et~al.\ for irregulars in the CfA redshift survey
\label{fig:loglf} }

\end{figure*}

Figure~\ref{fig:loglf} shows the luminosity function for the LSB
galaxies ($\mu(0) \geq 22.0$) from the APM survey.  The solid line
represents the maximum likelihood Schechter function from the STY
method, and the points with error bars represent the model-independent
SWML method.  As is obvious from Figure~\ref{fig:loglf}, the maximum
likelihood Schechter function is a very poor representation of the
``true'' distribution as determined by the SWML method: the reduced
\chisqnu\ from the likelihood ratio test of \cite{efstathiou88}
is 14.04, which implies that the probability of exceeding this
\chisqnu\ by chance is $\sim 2.5 \times 10^{-18}$.  The Schechter
function is particularly poor at the low-luminosity end.  There, the
model-independent SWML method finds two to three times the galaxy
density predicted by the maximum likelihood Schechter function.  The
SWML bins at $M = -16$, $-15$, $-14$, and $-13$ contain 31, 9, 15, and
12 galaxies respectively.  Across those four bins, the median
correction due to the APM selection function (Paper~II) is 0.759, and
the median correction due to the incomplete spectroscopic observations
(Equations~\ref{eqn:reasobs} and \ref{eqn:muasobs}) is 0.259; the
median total incompleteness correction (Equation~\ref{eqn:correct})
is therefore 0.197.  Our sampling of galaxies in these low-luminosity
bins is quite sparse, and hence the uncertainties at this end are
large.  The formal uncertainties shown in Figure~\ref{fig:loglf} may
well understate the true range.  

\begin{figure*}[t]
\epsfbox{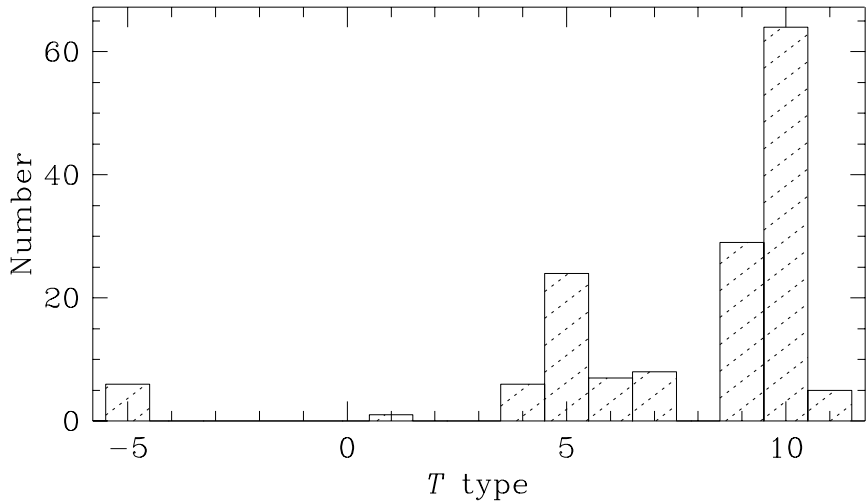}

\caption[Histogram of $T$-Types]{Histogram of $T$-types for LSB
galaxies from the APM survey.  Galaxies that appeared to be spirals
but whose images on the APM scans were too small to permit reliable
further classification were assigned $T=5$, so the number in that bin
is somewhat inflated.  Interacting galaxies were assigned $T=11$.
\label{fig:thist} }

\end{figure*}

The dashed line in Figure~\ref{fig:loglf} represents the Schechter
function estimated by \cite{marzke94a} for all galaxy morphologies in
the CfA Redshift Survey.  At the faintest luminosities, LSB galaxies
in the range $22.0 \leq \mu(0) \leq 25.0$ are more numerous than the
HSB galaxies sampled by the CfA Survey if the comparison is based on
the model-independent SWML points for the LSB galaxies, or
approximately as numerous if the comparison is based on the maximum
likelihood Schechter function.  For all galaxies brighter than $M_B <
-15$ the HSB galaxies are significantly more numerous.
Table~\ref{tab:schpars} lists the maximum likelihood Schechter
function parameters for the LSB galaxies in the present study, along
with similar parameters for all morphological types and for irregular
galaxies from the CfA Survey.

The LSB sample from the APM survey is not restricted as to
morphological type.  It includes a few dwarf ellipticals and early
spiral types.  However, it is dominated by very late-type spirals and
irregulars.  As Figure~\ref{fig:thist} shows, over half the LSB sample
have de~Vaucouleurs $T$-types of 9 or 10. For that reason, we have
also shown the Schechter function derived by \cite{marzke94a} for
irregulars (which they define as $8 \leq T \leq 10$) in
Figure~\ref{fig:loglf} and Table~\ref{tab:schpars}.  The Schechter
function for the CfA irregulars bears a striking resemblance to that
derived here for the LSB galaxies.  The steep low luminosity tail of
the function for CfA irregulars seems to match the model independent
SWML points for the LSB galaxies quite well.  This similarity in LF
slopes could be used as an argument that the high space density of
star forming irregulars are the parent population of the fainter LSBs.
Unfortunately, photometric surveys of LSBs continue to find no
relation between SB and color which is required to support such a
fading model.

\begin{table*}[htb]
\singlespace
\begin{center}
\caption{Comparison of Luminosity Function Model Parameters\label{tab:schpars}}
\begin{tabular}{lrrr}\\[0.2ex]
\hline\hline\relax\\[-1.7ex]
%
\multicolumn{1}{c}{Model/Survey} &
\multicolumn{1}{c}{$\alpha$} &
\multicolumn{1}{c}{$M^{*}$} &
\multicolumn{1}{c}{$\phi^{*}$} \\[0.2ex]
\multicolumn{1}{c}{(1)} &
\multicolumn{1}{c}{(2)} &
\multicolumn{1}{c}{(3)} &
\multicolumn{1}{c}{(4)} \\[0.7ex]
\hline\relax\\[-1.5ex]
%
 Maximum-Likelihood Schechter Functions:\\
 LSB & $-1.46$ & $-18.66$ & $0.0036$ \\
 CfA (all types) & $-1.02$ & $-18.90$ & $0.0201$ \\
 CfA (Sm - Im)   & $-1.87$ & $-18.79$ & $0.0006$ \\[1.5ex]
 Schechter Function $+$ Power Law:\\
 LSB (giants)    & $-0.92$ & $-18.19$ & $0.0060$ \\
 LSB (dwarfs)    & $-2.20$ & $-16.00$ & $0.0041$ \\[1.5ex]
\hline
\end{tabular}\\[1.5ex]
\end{center}
Notes: $M^{*}$ in $B$ mag, $\phi^{*}$ in \ihmpccmag.\\
\end{table*}

Several authors have suggested that the LF for faint galaxies may
exhibit an upturn from the pure Schechter form at faint luminosities.
The LF of \cite{impey88} for LSB dE's in the Virgo cluster turns up at
an apparent magnitude $m_B = 17$; the increase is so steep that they
were unable to rule out a divergent faint-end slope (\ie
$\alpha = -2.0$).  Upturns from the Schechter form have been observed
in Coma by \cite{thompson93}; in nearby local groups by
\cite{ferguson91}; in four local Abell clusters by \cite{depropris95};
and in Coma, Abell~2554 and Abell~963 by \cite{driver96b}.  These
deviations from the Schechter form are generally seen to begin in the
range $-17 \leq M_B \leq -15$, after adjustment to the distance scale
used here (\hubble).  In contrast, \cite{ferguson88} found an LF for
the Fornax cluster that was consistent with a single Schechter
function having a faint-end slope of $\alpha = -1.34$.  We note that
the form of the SWML data points in Figure~\ref{fig:loglf} is
consistent with earlier findings of a sharp change in slope at faint
luminosities: the binned model-independent data points clearly break
up from a smooth Schechter form at $M = -16$.

\begin{figure*}[t]
\epsfbox{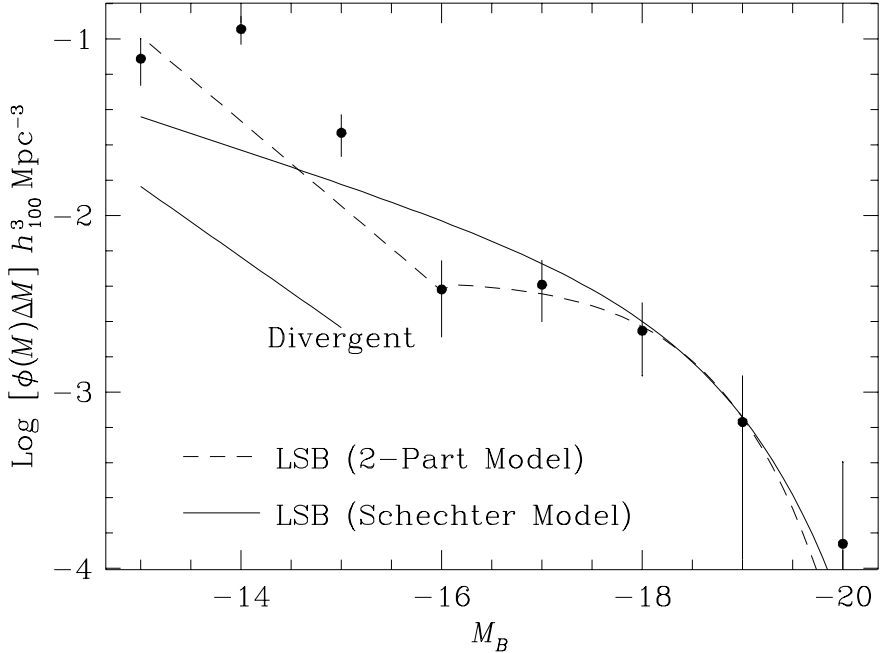}

\caption[Two-component luminosity function for LSB
galaxies]{Luminosity function for LSB galaxies from the APM survey,
determined using the two-component model described in the text (dashed
curve), and maximum likelihood Schechter function (solid curve).  The
points with error bars represent the model-independent step-wise
maximum likelihood LF.  For galaxies brighter than $M = -16$ the
two-component model is a standard Schechter function with all three
parameters allowed to vary.  For galaxies fainter than $M= -16$, the
model is a power law with the normalization at $M = -16$ constrained
to match the value of the Schechter component there.  Note that this
model was fit to the SWML binned LF, and not to the individual galaxy
data points.  The fiducial line labeled ``Divergent'' illustrates the
faint end slope $\alpha = -2$ where the integral of the LF becomes
divergent. \label{fig:twocomp}}

\end{figure*}

To investigate this break in more detail, we also fit a two-component
model to the SWML LF representation.  For absolute magnitudes $M \le
-16$ (``giants''), the model followed the usual Schechter function
form, with all three parameters ($\alpha$, $M^{*}$, and $\phi^{*}$)
allowed to vary.  For absolute magnitudes $M \ge -16$ (``dwarfs''),
the model followed a simple power law, with the normalization at $M =
-16$ constrained to match the Schechter model value there (\ie only
the slope $\alpha$ was allowed to vary in the fit).  Results of this
fit are depicted in Figure~\ref{fig:twocomp}, and the fitted model
parameters are listed in Table~\ref{tab:schpars}.  Two aspects of this
fit deserve special comment.  First, it is not possible to compare
directly the goodness-of-fit of this two-component model to that of
the STY maximum-likelihood Schechter function.  The \chisqnu\ quoted
above for the STY result is obtained by a likelihood-ratio test which,
like the STY model itself, is computed from the individual galaxy data
points.  The two-component model is a fit to the binned SWML LF, not
to the individual galaxy data points, and thus its much larger
\chisqnu\ ($\chisqnu = 294.8$, due to the much smaller number of
degrees of freedom) is computed in a very different manner.  Second,
the two-component model yields results which are not physically
plausible.  The resulting slope of the dwarf galaxy power law is
$\alpha = -2.20$, which implies an infinite total luminosity if the
two-component LF is integrated from zero to infinity.  This faint-end
slope is best interpreted as a finding that the model-independent SWML
indicates a very steeply increasing density of faint galaxies, so
steep in fact that a divergent LF cannot be ruled out (\cf Impey,
Bothun \& Malin 1988\nocite{impey88}).

\begin{figure*}[t]
\epsfbox{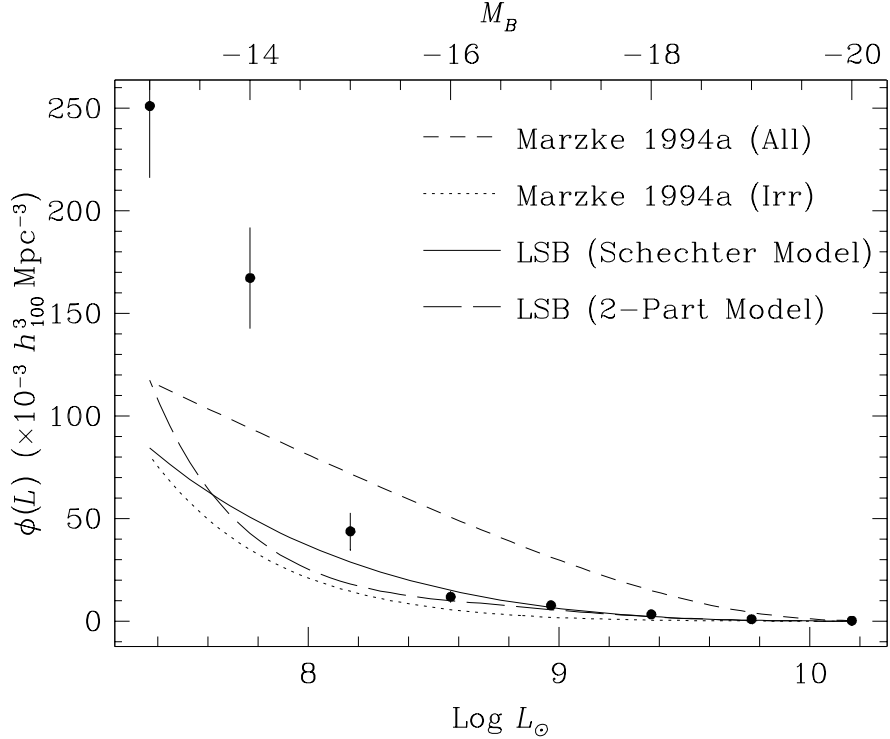}

\caption[Cumulative luminosity function for LSB galaxies]{Cumulative
luminosity functions for LSB galaxies and for galaxies from the CfA
redshift survey, plotted linearly as a function of Log $L$.  The top
axis represents the corresponding $B$ magnitude.  The cumulation runs
from high to low luminosities (\ie from right to left).
The points represent the binned model independent LF for the LSB
galaxies, the solid line represents the maximum likelihood
Schechter function for the LSB galaxies, and the long-dashed line
represents the two-component model LF for the LSB galaxies.  The
short-dashed line is the Schechter function determined by Marzke
et~al.\ (1994) for all morphological types in the CfA redshift survey,
and the dotted line is the Schechter function determined by Marzke
et~al.\ (1994) for irregulars. \label{fig:cnormlin} }

\end{figure*}

Because the low luminosity tail of the LSB luminosity function rises
so steeply, the contribution of LSB galaxies to the overall number
density of field galaxies locally is quite large.
Figure~\ref{fig:cnormlin} shows the same luminosity functions as
Figure~\ref{fig:loglf} along with the two-component LF described
above, but cumulated to show total number densities.  The cumulation
runs from high to low luminosities (\ie from right to
left), and the vertical axis scaling is linear.  Across the whole
range of luminosities, the LSB galaxies are almost twice as numerous
as all the HSB galaxies in the CfA survey if the comparison is based
on the SWML points for the LSB galaxies, or half as numerous if the
comparison is based on the maximum likelihood Schechter function.
Thus, even under the most conservative estimate, surveys like the CfA
redshift survey have missed at least one-third of the local galaxy
population due to surface brightness selection biases.  The true
missed fraction is almost certainly higher, even by the most
conservative estimator.  The LSB LF presented here covers only the
range $22.0 < \mu(0) \la 25.0$, but as \cite{mcgaugh95a} and Paper~II
showed, the distribution appears flat for SB levels $\mu(0) \geq
25.0$.  Thus, surveys sensitive to fainter SB levels should find even
higher number densities of galaxies.

\begin{figure*}[t]
\epsfbox{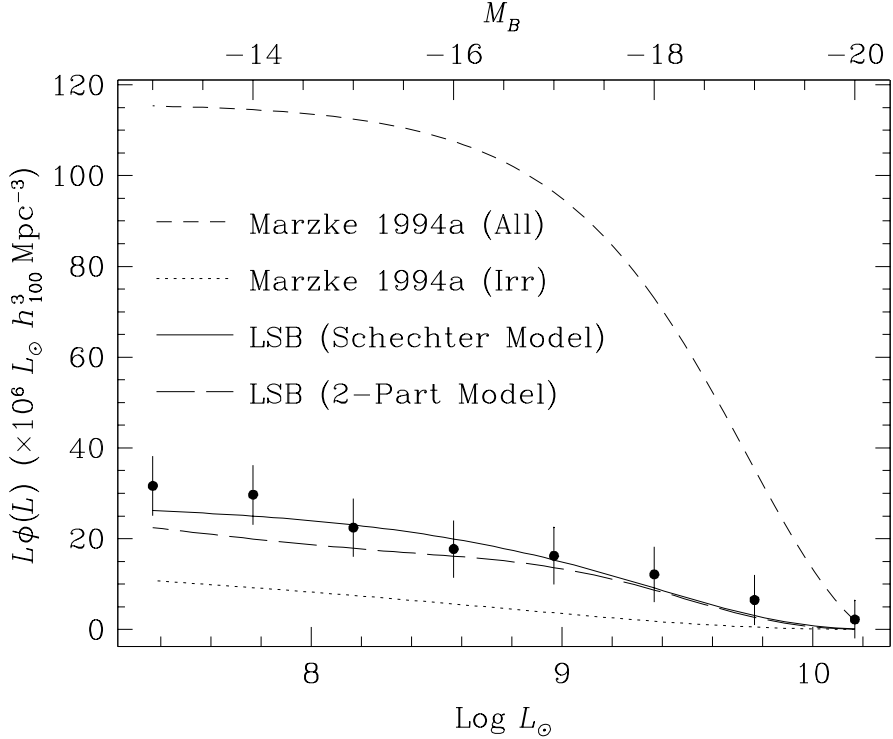}

\caption[Cumulative luminosity densities]{Cumulative luminosity
densities for LSB galaxies and for galaxies from the CfA redshift
survey, plotted linearly as a function of Log $L$.  The top axis
indicates the corresponding $B$ magnitude.  The points represent the
binned model independent LF for the LSB galaxies, the solid represents
the maximum likelihood Schechter function for the LSB galaxies, and
the long-dashed line represents the two-component model LF for the LSB
galaxies.  The short-dashed line is the Schechter function for all
morphological types in the CfA redshift survey determined by Marzke
et~al.\ (1994) and the dotted line is the Schechter function
determined by Marzke et~al.\ (1994) for irregulars.  The cumulation
runs from high to low luminosities (\ie from right to
left). \label{fig:clumlin} }

\end{figure*}

Despite the significance of their total numbers, LSB galaxies
contribute little to the total luminosity density of the local
universe, because the highest number densities of LSB galaxies occur
at the lowest galaxy luminosities.  Figure~\ref{fig:clumlin} shows the
cumulative luminosity densities for LSB galaxies from the APM survey
and HSB galaxies from the CfA redshift survey.  Again, the cumulation
runs from high to low luminosities (right to left) and the vertical
scaling is linear.  By any estimator, LSB galaxies represent a small
fraction of the total luminosity emitted by HSB galaxies from the CfA
survey.

\section{Implications} \label{sec:imp}

Much attention has been devoted over the past 15 years to the
population of faint blue galaxies revealed in surveys sensitive to
extended objects as faint as $m_{B_J} \sim 27$.  These galaxies become
bluer at fainter apparent magnitudes (Lilly \etal
1995\nocite{lilly95c}).  They generally are not at extreme redshifts:
\cite{lilly95c} found a median redshift of $z_{med} \approx 0.56$ for a
sample of galaxies in the magnitude range $17.5 < I_{AB} < 22.5$.
These galaxies are clustered more weakly than are most local bright
galaxies, though their clustering strength is roughly comparable to
that of local galaxies undergoing rapid star formation, per
\cite{bernstein94a}.  Their numbers are significantly in excess of
expectations based on local galaxy populations in the absence of
evolution (Tyson 1988\nocite{tyson88}; Lilly et~al.\
1991\nocite{lilly91}; McLeod \& Rieke 1995\nocite{mcleod95}).  This
excess has led some authors to suggest non-standard cosmologies as a
possible explanation (Yoshii 1993\nocite{yoshii93}), and others to
propose strong evolution in galaxy luminosities, perhaps with the rate
of evolution itself a function of luminosity (Broadhurst et~al.\
1988\nocite{broadhurst88}; Babul \& Rees 1992\nocite{babul92}; Babul
\& Ferguson 1996\nocite{babul96}).  Still another approach, taken by
\cite{gronwall95}, is to derive local luminosity functions by finding
functions that can explain as well as possible the faint galaxy number
counts without invoking strong evolution.  The luminosity functions
they derive predict more local low-luminosity galaxies than are
observed in existing surveys.  At the very least, recent surveys for
LSB galaxies indicate that the galaxy density at $z=0$ is higher than
previously assumed which means, at some level, the apparent excess of
faint galaxies at high redshifts is at least in part an artifact of
improper normalization at $z=0$.  The issue is how large this effect
really is.

\cite{mcgaugh94b} suggested that LSB galaxies such as those in the
present sample could help reconcile the differences between observed
local populations and this population of faint blue galaxies
(hereafter, FBGs).  He noted that, like the FBGs, LSB galaxies are
generally blue (McGaugh \& Bothun 1994\nocite{mcgaugh94}) and weakly
clustered (Mo et~al.\ 1994\nocite{mo94}).  Furthermore, if current
models of slow, continuous star formation LSB galaxies are correct
(McGaugh \& Bothun 1994\nocite{mcgaugh94}), \cite{mcgaugh94b} argued
that LSB galaxies should become only slightly redder over the
timescales of interest, $0 < z \la 0.5$.  He also demonstrated through
a simple analytic calculation that the deep CCD surveys would be more
sensitive to LSB galaxies at $z \sim 0.4$ than wide-field photographic
surveys are to local ($z \la 0.1$) LSB galaxies.  He argued that
including nearby LSB galaxies in the local luminosity function could
reconcile the number of low-luminosity galaxies in the local
population with the FBG population.  

More recently still, \cite{driver95} and \cite{driver95a} have
examined the morphological mix of the faint field galaxies using data
from the Hubble Space Telescope, down to a flux limit of $m_I =
24.25$.  They compared the observed differential number counts (number
per square degree as a function of apparent magnitude) for three
different morphological groupings to the predictions of various
models.  They concluded that differential number counts of ellipticals
and early-type spirals are consistent with the predictions of a
no-evolution model based on standard local LFs (after renormalization)
for these types taken from \cite{marzke94a} and \cite{loveday92}.
They also found that the observed number counts of late-type
spirals and irregulars were substantially in excess of similar
no-evolution predictions for these classes.  They could reconcile
prediction with observation for this morphological class only by
including a substantial amount of luminosity evolution (1.3 magnitudes
of brightening by $z \sim 0.5$ for an Irr LF from Marzke
et~al.\ (1994)) or by using a sharply increased normalization
($\phi^{*} = 3.5 \times 10^{-2}\, \ihmpcc$; compare with
Table~\ref{tab:schpars}) for the Irr LF.  

Our results underscore the uncertainty in the faint end slope of the
field galaxy luminosity function.  \cite{driver96b} have shown that
existing wide field redshift surveys place few constraints on the
shape of the luminosity function below $M_B = -16.5$ assuming
\hubble. By reaching lower in surface brightness, we have isolated a
population of blue, LSB dwarfs that is absent from published
luminosity functions, and which contributes strongly below $M_B =
-16$.  \cite{marzke94a} also saw evidence for a sharp upturn in the
dwarf and irregular population at about the same luminosity. More
recently, \cite{zucca96} have shown that the luminosity function of
\cite{loveday92} calculated from the Stromlo-APM survey is
significantly incomplete. The new ESO Slice luminosity function shows
an upturn below $M_B = -16$, due to blue, star-forming galaxies, made
up of a mixture of compact dwarfs and LSB galaxies. In all these
studies, as in the deeper HST surveys of \cite{driver95a}, the data
are well described by a hybrid luminosity function consisting of a
bright end Schechter function with $\alpha = -1$, and a faint end
($M_B \ga -16$ assuming \hubble) power law with a slope $-1.4 < \alpha
< -1.8$. These faint galaxies are not major contributors to the
luminosity density of the universe, but because the trend of $M/L$
with luminosity is not well understood their contribution to the mass
density is an open issue.

There is also now evidence that some evolution may have occurred in the
late-type galaxy population over the range $0 \leq z \leq 1$.
\cite{lilly95c} have studied the evolution of the LFs over this range
using date from the recent Canada-France Redshift Survey (CFRS).  They
found that the LF for red galaxies shows little change in number
density or luminosity over this range in $z$, but that the LF for blue
galaxies appears to have brightened uniformly by about 1 magnitude by
$z \sim 0.75$.  

Thus, the ``excess'' of FBGs consists of late-type spirals and
irregulars, the same types which dominate the population of LSB
galaxies found by the APM survey (Fig~\ref{fig:thist}).  These LSB
galaxies have a Schechter function normalization approximately 6 times
as large as that found by \cite{marzke94a} for HSB irregulars, so they
expand the known local population well beyond that used by
\cite{driver95} in their modeling.  The LSB normalization still is not
as large as that found necessary by \cite{driver95} to account for the
FBG counts with no evolution, but the heretofore-uncounted LSB
galaxies do help considerably to close the gap between local
population estimates and the FBG counts.  Taking this increase
together with the modest evolution observed in blue galaxy LFs by
\cite{lilly95c}, it may now be possible to make an essentially
complete reconciliation between local populations and the FBGs.
However, any such reconciliation will also require a model for the
evolution of the FBGs which accounts for the very blue colors of both
the FBGs and the local LSBs (see McGaugh \& Bothun
1994\nocite{mcgaugh94}, McGaugh 1994\nocite{mcgaugh94b}, and McGaugh
et~al.\ 1995a\nocite{mcgaugh95}).  

We can demonstrate the rough equivalence between the local LSB dwarf
population and the MDS population of \cite{driver95a}, which we
presume to be at typical redshifts $0.3 \la z \la 0.6$ based on
\cite{lilly95c}. The LSB dwarfs with $M_B > -16$ are at typical
distances of $10 \la d \la 40$ \hmpc. They have central surface
brightness in the range $22.0 < \mu_B(0) \la 25$ \magsq, and effective
angular radii of $6 \la r_{eff} \la 20$ arcseconds. If they are related to
the LSB dwarfs in clusters, we expect them to have $\bv \sim 0.5$
(Impey, Bothun, \& Malin 1988\nocite{impey88}). The late-type and
irregular (Sdm/Irr) MDS galaxies have effective radii with a median
value of 0.4 arcseconds (Im et~al.\ 1995\nocite{im95}), which would scale
to $20 \la r_{eff} \la 40$ arcseconds for a local population.  The central
surface brightnesses of disk-dominated categories in the MDS sample
are $20 \la \mu_I(0) \la 22$ \magsq per \cite{mutz94}, which is
equivalent to $22 \la \mu_B(0) \la 24$ \magsq\ locally, assuming no
evolution. The Sdm/Irr galaxies have observed colors of $\vi \sim 1$
per \cite{casertano95},
consistent with a local star-forming dwarf color of $\bv \sim 0.5$,
once again assuming no evolution. There could well be a subset of the
MDS population which fade in surface brightness and redden to below
the detection threshold for our blue photographic survey. Such
galaxies would be absent from all local catalogs.

\section{Conclusions} \label{sec:cnc}

We have estimated a luminosity function for galaxies with surface
brightnesses fainter than $\mu(0) = 22.0 \, \magsq$, which is the
approximate faint limit of $\mu(0)$ for galaxies covered by the CfA
Redshift Survey.  We find that this LSB LF has a steeply rising tail
at low luminosities($\alpha = 1.42$), comparable to that found by
\cite{marzke94a} for galaxy types $8 \leq T \leq 10$.  The LSB LF has
a normalization lower than that found for the overall CfA survey,
but much higher than that found for types $8 \leq T \leq 10$.  Thus
estimates of the total population of local galaxies based on the CfA
survey are missing at least one-third of the total number of galaxies
due to surface brightness selection bias.  These previously
unaccounted-for LSB galaxies can help considerably to resolve the
apparent difference between estimates of the local population and the
large numbers of faint blue galaxies observed at moderate redshift.

\acknowledgments

We are grateful to a number of our colleagues for many stimulating and
helpful discussions.  We thank in particular: Frank Briggs, Erwin de
Blok, Simon Driver, Marijn Franx, Ron Marzke, Stacy McGaugh, Renzo
Sancisi and Martin Zwaan.  We also thank the staffs of the Multiple
Mirror Telescope Observatory, the Steward Observatory Kitt Peak
Station, and the Arecibo Observatory for their expert assistance
during the many observing runs carried out in connection with our
survey.  This project made extensive use of the NASA Astrophysics Data
System.  This work was supported in part by the
National Science Foundation under Grant AST-9003158.  

\clearpage

\setlength{\parsep}{0ex}\huge\normalsize

\bibliography{astr1990,astr1991,astr1992,astr1993,astr1994,astr1995,astr1996}

\end{document}